\documentclass[aps,showpacs,%amsmath,
amssymb,endfloats]{revtex4}
\usepackage{amsmath}
\usepackage{bm,amsfonts,graphicx}
\begin{document}
\title {Controlling statistical properties of stored light}
\author{A. Raczy\'nski}
\email{raczyn@fizyka.umk.pl}
\author{K. S\l owik}
\author{J. Zaremba}
\affiliation{Instytut Fizyki, Uniwersytet Miko\l aja Kopernika, ulica Grudzi\c{a}dzka 5,
87-100 Toru\'n, Poland,}
\author{S. Zieli\'nska-Kaniasty}
\affiliation{Instytut Matematyki i Fizyki, Uniwersytet Technologiczno-Przyrodniczy, Aleja
Prof. S. Kaliskiego 7, 85-796 Bydgoszcz, Poland.}
\begin{abstract}
Statistical properties of outgoing light pulses are studies after they have been stored in
a medium of atoms in the tripod configuration. A generalized Hong-Ou-Mandel interference,
storing of squeezed states and homodyne signal analysis are discussed in the context of
their dependence on the parameters of the control fields used for light storage and
release. 
\end{abstract}
\maketitle
\newpage
%\section{Introduction}
Recent intensive studies of light slowdown and storage in optically driven atomic systems
have been motivated both by their fundamental aspects and potential applications. The
minimum model system in which such processes can occur includes an atomic medium of
three-level atoms irradiated by two laser beams (signal and control one) in the $\Lambda$
configuration. The underlying physical scenario consists in modifying electromagnetically
induced transparency (EIT) in time and, as a consequence, in a reversible mapping of the
laser signal onto the atomic excitation represented by a ground state atomic coherence;
for recent reviews see, e.g., Refs. \cite{lukin,andre,imamoglu}. A more advanced system is
a medium of four-level atoms interacting with three laser fields in the tripod
configuration. The essential new element is that more atomic coherences become relevant
and one has more external parameters to control the processes. One of the possibilities of
exploring such a system is a situation in which two control fields are used during the
first stage of the storage procedure to map a signal onto a certain superposition of two
atomic coherences. The form of this superposition is determined by the parameters of the
control fields. Due to the possibility of creating the other superposition of the
coherences, orthogonal to the former one, another pulse can be stored at the following
stage, without affecting the excitation created at the first stage. Similarly the stored
excitation can be released in two steps, in general from different orthogonal
superpositions of the coherences. This means that our system is capable of mixing two
signals separated in time and of releasing the signals at two different time instants
(signal mixing and/or splitting). One can thus make two time-separated pulses interfere at
the storage stage and study statistical properties of the retrieved signals. Note that the
situation, in which there is only a single stored signal in the classical description, in
a complete quantum analysis requires treating the other signal as "incoming vacuum" which
does influence the outgoing signals. The discussed system can thus be considered a kind of
a beam splitter or interferometer in the time domain, operating on stored light, and the
above-mentioned two consecutive stages of storing (releasing) light can be considered
input (output) ports. Moreover, additional manipulations on atomic coherences are possible
at the storage stage. An essential difference compared with standard devices is that four
light beams (two incoming and two outgoing ones) propagate in the same direction but are
well separated in time. In contradistinction to usual beam splitters here it is possible
to smoothly regulate amplitude and phase relations of the outgoing light (as if one could
change on demand the transmission of the splitter and the phase jump of the reflected
light). With such a flexible analogue of a beam splitter it is reasonable and interesting
to check how statistical properties of the outgoing light depend on the parameters of the
control fields and/or additional interactions at the storage stage. Note that some
statistical properties of photon states transformed by a usual beam splitter have been
discussed in Ref. \cite{campos}

A medium of atoms in the tripod configuration has already been discussed in the context of
light propagation. Paspalakis and Knight \cite{pasp} studied propagation and slowdown of a
weak classical probe pulse and discussed the susceptibility of the medium dressed by two
control fields.  Wang {\em et al.} \cite{yelin} theoretically performed splitting of a
single photon by storing it in one atomic coherence, by transferring a part of the
excitation to the other one by F-STIRAP technique and finally by releasing the pulse from
the latter. Mazets investigated a simultaneous adiabatic propagation of three laser beams
of comparable intensities \cite{mazets1} and has analyzed propagation of two quantum
signals propagating in the tripod medium dressed by a single control field and
demonstrated the existence of two modes propagating with different velocities
\cite{mazets2}. Petrosyan and Malakyan \cite{petrosyan} considered propagation of two
orthogonal polarization components of a single beam and have studied the effect of
nonlinear interaction of the two probes. In our earlier paper \cite{my1} we have studied
storing of a classical light pulse in the tripod medium driven by two control fields.

An important class of questions are those concerning quantum statistical properties of
light pulses, not only because of their fundamental physical aspect but also in the
context of information storage and retrieving. In Ref. \cite{my2} we have investigated
quantum interference of photons in this context and we have shown how to realize the
Hong-Ou-Mandel \cite{hong, knight} interference in such a system. We have also discussed
how certain statistical properties of the stored light are modified by vacuum
fluctuations. The aim of the present paper is to generalize those investigations in the
case in which the input light states are various non-classical states, the analyzed
quantities are not only the photon number but also quadrature operators and the homodyne
signal. In particular we study the dependence of the statistical light properties on the
parameters of the control fields.

%\section{Polaritons}
The atomic medium considered is composed of four-level atoms with an upper state $a$ and
three lowers states  $b$ (initial), $c$ and $d$. The quantum signal field 1 couples the
states $b$ and $a$ while two classical control fields 2 and 3 couple the state $a$
respectively with $c$ and $d$.

The realization of beam splitting/mixing consists in storing two light pulses in two
consecutive time-separated steps by switching off at the same rate two sets of the control
fields of precisely chosen amplitudes (slowly varying at the same rate) and phases
(assumed constant); for details see ref. \cite{my2}. Switching the control fields off
makes the medium opaque and the signals are stored independently in coherent
superpositions of the atomic coherences. In the first step the Rabi frequencies and phases
of the control fields are arbitrary, say, respectively, $\Omega_{2}^{0}$,
$\Omega_{3}^{0}$, $\chi_{2}^{0}$ and $\chi_{3}^{0}$, with the mixing angle is defined as
$\phi^{0}=\arctan\frac{\Omega_{3}^{0}}{\Omega_{2}^{0}}=const$. In the second step the
corresponding control fields are characterized by the angles $(\frac{\pi}{2}-\phi_{0}$,
$\chi_{2}^{0}+\pi$, $\chi_{3}^{0}$). As a consequence, two signals with shapes given by
the wave packets $f_{1}(z)$ and $f_{2}(z)$ of overlap $s=\int f_{1}^{*}(z)f_{2}(z) dz$ are
independently stored in two orthogonal combinations of the atomic coherences $\sigma_{bc}$
and $\sigma_{bd}$. Releasing the pulses takes places again in two time-separated steps by
switching on at the same rate two sets of the control fields characterized, respectively,
by arbitrary angles $(\phi^{1}$, $\chi_{2}^{1}$, $\chi_{3}^{1})$ and
($\frac{\pi}{2}-\phi^{1}$, $\chi_{2}^{1}+\pi$, $\chi_{3}^{1}$). The relations between the
amplitudes and phases of the control fields at the storing and release stages determine
how, i.e. with which amplitudes and phases, the stored excitation is divided into two
time-separated outgoing signals. An additional control procedure is also possible
consisting in modifying the atomic coherences at the storage stage by some additional
interaction.

We assume that the evolution of our system is adiabatic and can be described
perturbatively with respect to the signal field, resonance conditions are fulfilled and
relaxation processes and noise effects can be neglected. In fact it has been proven by
Peng {\em et al.} \cite{peng} that such quantum properties like squeezing and entanglement
do not suffer much due to dephasing and noise in EIT $\Lambda$ medium. Note however that
this observation may not be true in general, e.g., for the two beams being of comparable
intensities or far from the two-photon resonance \cite{barberis}. Under such assumptions
the evolution can be described in terms of joint atom+field excitations called polaritons
which are described by photon and/or atomic excitation annihilation operators
\begin{eqnarray}
&X^{0}_{1}(z,t)=\frac{1}{g\sqrt{L}}\{\epsilon^{(+)}(z,t)
\cos\theta-\frac{\hbar\kappa}{d_{ab}}
\sin\theta[\exp(i\chi_{2}^{0})\cos\phi^{0}\sigma_{bc}(z,t)+
\exp(i\chi_{3}^{0})\sin\phi^{0}\sigma_{bd}(z,t)]\},
\nonumber\\
&X^{0}_{2}(z,t)=\frac{1}{g\sqrt{L}}\{\frac{\hbar\kappa}{d_{ab}}
[\exp(i\chi_{2}^{0})\sin\phi^{0}\sigma_{bc}(z,t)-\exp(i\chi_{3}^{0})
\cos\phi^{0}\sigma_{bd}(z,t)]\}.
\end{eqnarray}
In the above formula $\epsilon^{(+)}$ is the positive frequency part of the signal
electric field, $d$ is the dipole moment operator,
$\sigma_{\alpha\beta}=|\alpha><\beta|\exp(ik_{\alpha\beta}z)$ are atomic flip operators,
$g=\sqrt{\frac{\hbar\omega}{2\epsilon_{0}V}}$,
$\kappa^{2}=N|d_{ab}|^{2}\omega/(2\epsilon_{0}\hbar)$, $N$ is the atom density, $V$ is the
quantization volume, $L$ is the length of the sample, $\omega$ is the field frequency,
$k_{\alpha\beta}=(E_{\alpha}-E_{\beta})/(\hbar c)$,
$\tan^{2}\theta=\kappa^{2}/[(\Omega_{2}^{0})^{2}+(\Omega_{3}^{0})^{2}]$. Changing the
control fields corresponding to the first stage of storing (release) to those
corresponding to the second stage means in fact an interchange of the polaritons $X_{1}$
and $X_{2}$.

Thus an annihilation operator of a photon stored at the first (second) stage as a packet
$f_{1}$ ($f_{2}$) becomes an atomic annihilation operator $X^{0}_{1}(1)$ ($X^{0}_{2}(2)$),
obtained by expanding the operators of Eq. (1) in the two-element basis (in our previous
paper \cite{my2} we have denoted $X_{1}$ by $\Psi$ and $X_{2}$ by $Z$). In an analogous
way the operators of the outgoing photons, before they have been released, are atomic
excitation operators $X^{1}_{1,2}(1,2)$. The main idea of the theoretical description is
based on representing each polariton operator at the release stage by a superposition of
operators at the storage stage (or vice versa)
\begin{equation}
X_{j}^{1}(m)=\sum_{k=1,2} S_{jk} X_{k}^{0}(m),
\end{equation}
where $j=1,2$, $m=1,2$ and the unitary matrix $S$
\begin{eqnarray}
S_{11}=\cos\phi^{1}\cos\phi^{0}\exp[i(\chi_{2}^{1}-\chi_{2}^{0})] +
\sin\phi^{1}\sin\phi^{0}\exp[i(\chi_{3}^{1}-\chi_{3}^{0})],\nonumber\\
S_{12}=-\cos\phi^{1}\sin\phi^{0}\exp[i(\chi_{2}^{1}-\chi_{2}^{0})]
+\sin\phi^{1}\cos\phi^{0}\exp[i(\chi_{3}^{1}-\chi_{3}^{0})],\nonumber\\
S_{21}=-\sin\phi^{1}\cos\phi^{0}\exp[i(\chi_{2}^{1}-\chi_{2}^{0})] +
\cos\phi^{1}\sin\phi^{0}\exp[i\chi_{3}^{1}-\chi_{3}^{0})],\nonumber\\
S_{22}=\sin\phi^{1}\sin\phi^{0}\exp[i(\chi_{2}^{1}-\chi_{2}^{0})] +
\cos\phi^{1}\cos\phi^{0}\exp[i\chi_{3}^{1}-\chi_{3}^{0})].\nonumber\\
\end{eqnarray}
The commutation relations pertain in general to nonorthogonal modes (due to an overlap of
the wave packets $f_{1,2}$) and read
\begin{equation}
[X_{j}^{p}(m),X_{k}^{p\dagger}(n)]=\delta_{jk}\gamma_{mn},
\end{equation}
where $j,k,m,n =1,2$ , $p=0,1$ and
\begin{equation}
\gamma=\left(\begin{array}{cc}1&s\\s^{*}&1\\ \end{array}\right).
\end{equation}

%\section{Examples}
This formalism allows one to investigate how quantum statistical properties of the
released light are modified due to its storage, depending on the characteristics of the
control fields and possibly on phase operations on the atomic coherences into which the
pulse has been mapped.

Consider first the case of quantum interference of multiple photons \cite{mandel}. The
incoming photons states are then definite Fock states. Note that the cases of interference
of three and four photons on standard, possibly non-symmetric beam splitters have recently
been studied both theoretically and experimentally \cite{wang, liu}.

Storing $n$ and $m$ photons at, respectively, first and second storing stage means that
the medium is in the quantum state
\begin{equation}
|\psi>=\frac{1}{\sqrt{n!m!}}(X_{1}^{0\dagger}(1))^{n} (X_{2}^{0\dagger}(2))^{m}|0>,
\end{equation}
where the "vacuum" $|0>$ denotes the state in which all the medium atoms are in the ground
state. Expressing the operators $X_{j}^{0}$ by $X_{k}^{1}$ (Eq. (2)) one gets
\begin{equation}
|\psi>\equiv\sum_{i}|\psi_{i}>=\sum_{i}\frac{1}{\sqrt{n!m!}}\sum_{k}\binom{n}{k}\binom{m}{i-k}
S_{11}^{k}S_{21}^{n-k}
S_{12}^{i-k}S_{22}^{m-i+k}[X_{1}^{1\dagger}(1)]^{k}[X_{2}^{1\dagger}(1)]^{n-k}
[X_{1}^{1\dagger}(2)]^{i-k}[X_{2}^{1\dagger}(2)]^{m-i+k} |0>.
\end{equation}
Each component $\psi_{i}$ corresponds to releasing $i$ photons at the first release stage
and $n+m-i$ photons at the second one, irrespective from the spatial localization of the
excitation. The norm $<\psi_{i}|\psi_{i}>=P(i)$ gives thus the probability of such a
photon final distribution, which can be easily calculated when storing at the two stages
occurred at the same localization and with the same profile $f_{1}=f_{2}$ (i.e. $s=1$; in
such a case we drop the arguments 1 and 2 of $X$)
\begin{equation}
P(i)=\frac{i!(n+m-i)!}{n!m!}\sum_{kk'}\binom{n}{k}\binom{n}{k'}\binom{m}{i-k}\binom{m}{i-k'}
S_{11}^{k}S_{21}^{n-k}S_{12}^{i-k}S_{22}^{m-i+k}S_{11}^{*k'}S_{21}^{*n-k'}S_{12}^{*i-k'}
S_{22}^{*m-i+k'}.
\end{equation}
For non-overlapping packets inside the sample, i.e. for $s\neq1$, the calculations are
tedious and can be performed in a recursive way.

One can also get simple formulae for the mean value and variance of the final photon
distribution. The polariton number operator to describe photons released at the first
release stage in the case of arbitrary packet overlap $s$ is
\begin{equation}
N_{1}^{1}=\frac{1}{1-|s|^{2}}\sum_{jk}(-1)^{j+k}X_{1}^{1\dagger}(j)X_{1}^{1}(k)\gamma_{jk}.
\end{equation}
The latter formula can be obtained as a sum of numbers of excitations from
Schmidt-orthogonalized modes and reduces to the standard expressions:
$X_{1}^{\dagger}(1)X_{1}(1)$ for $s=1$ or $X_{1}^{\dagger}(1)X_{1}(1)+
X_{1}^{\dagger}(2)X_{1}(2)$ for $s=0$. An analogous formula holds for the photons released
at the second stage described by $N_{2}^{1}$. The mean values of $N_{1}^{1}$ and its
square can be calculated from Eqs. (9), (6) and (2) by a multiple usage of the commutation
relations to finally give
\begin{equation}
\overline{N_{1}^{1}}=|S_{11}|^{2}n+|S_{12}|^{2} m,
\end{equation}
and
\begin{equation}
\overline{[N_{1}^{1}]^{2}}= |S_{11}|^{4}n^{2}+|S_{12}|^{4}m^{2}
+|S_{11}|^{2}|S_{12}|^{2}[2nm(1+|s|^{2})+m+n].
\end{equation}

The Fano factor being the ratio of the variance $W$ and the corresponding mean value is
thus
\begin{equation}
\frac{W(N_{1}^{1})}{\overline{N_{1}^{1}}}=\frac{|S_{11}|^{2}|S_{12}|^{2}(2nm|s|^{2}+m+n)}
{|S_{11}|^{2}n+|S_{12}|^{2}m}.
\end{equation}
In particular for equal photon numbers in the incoming channels ($n=m$)
\begin{equation}
\frac{W(N_{1}^{1})}{\overline{N_{1}^{1}}}=2|S_{11}|^{2}|S_{12}|^{2}(n|s|^{2}+1).
\end{equation}
Eq. (12) is a generalization of our earlier results obtained in the case of vacuum in the
second input channel. Though the variances in both input channels are zero the variance in
the output channel 1 (as well as in the output channel 2) is nonzero. Thus the outgoing
states are no longer Fock states and the Fano factor, being here the measure of deviation
of the ouput state with respect to the input state, is maximum if the overlap $s=1$, i.e.
when both input signals have the same shape and have been stored at the same place inside
the sample. One can also see a continuous transition to the situation in which the input
signals do not overlap at all.

It can be seen from Eq. (1) that instead of changing the phase $\chi_{2,3}$ of the control
fields one can realize an equivalent modification of the polaritons by introducing
additional modifications of the atomic coherences $\sigma_{\alpha\beta}$. For example, as
we have mentioned in our previous paper on the propagation and storage of classical pulses
\cite{my1}, one can make such a choice of the atomic states that switching on a magnetic
field parallel to the propagation direction induces level shifts of the atomic levels and
thus a phase shift $\delta$ of the coherence $\sigma_{bc}$ while leaving $\sigma_{bd}$
unchanged. If, e.g., the states $b$, $c$, and $d$ are Zeeman sublevels of the hyperfine
structure with the quantum numbers $(F,M)$ taken respectively (2,-1), (2,1), and (1,1),
the coherence $\sigma_{bc}$ will acquire a phase shift of absolute value $\delta=eA/(2m)$,
where $A$ is the area of the magnetic pulse ($A=Bt$ for a rectangular pulse of induction
$B$ and duration $t$). Introducing the phase $\delta$ is equivalent to changing the phase
$\chi_{2}^{0}$. Thus if the amplitudes of the control fields 2 and 3 both at the storing
and release stages are equal, which means $\phi^{0}=\phi^{1}=\pi/4$ and the phases of the
control fields are zero, the polariton transformation matrix (Eq. (3)) has the elements
$S_{11}=S_{22}=\exp(-i\frac{\delta}{2})\cos\frac{\delta}{2}$, $
S_{12}=S_{21}=i\exp(-i\frac{\delta}{2})\sin\frac{\delta}{2}$. An experiment of the
Hong-Ou-Mandel type would then consist in trapping single photons ($m=n=1$) of the same
shape at the same place of the sample. The probability of releasing a single photon at the
first release stage ($i=1$) given by Eq. (8) is
\begin{equation}
P(1)=\cos^{2} \delta,
\end{equation}
which means that if no phase modification has been made, the two channels do not mix at
all and each photon is released as it has been trapped.The effect can be modulated by
changing the phase $\delta$ due to a magnetic pulse. Photon coalescence occurs for
$\delta=\frac{\pi}{2}$. Another typical realization of the standard HOM interference
occurs for $\chi_{i}^{j}=0$; in such a case $P(1)=\cos^{2}(\phi^{1}-\phi^{0})$.

\begin{figure}
\includegraphics{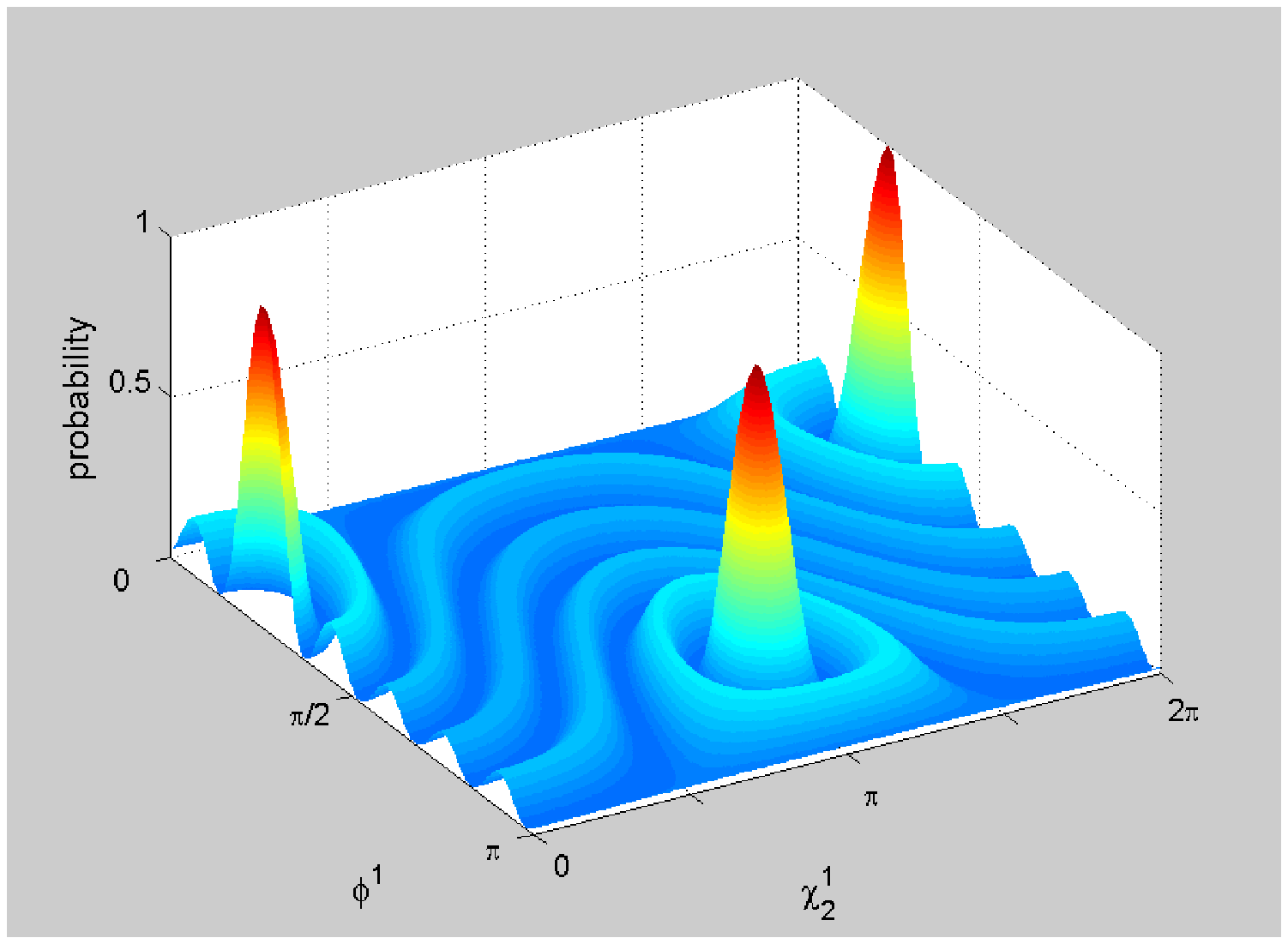}
\caption {\label{fig1} The probability of observing exactly 6 photons in the first output
channel in the case of 6 photons entering in both input channels as a function of the
angles $\phi^{1}$ and $\chi_{2}^{1}$ for $\phi^{0}=\pi/8$ and
$\chi_{2}^{0}=\chi_{3}^{0}=\chi_{3}^{1}=0$.}
\end{figure}

As a more general example we illustrate in Fig. \ref{fig1} the situation in which $n=m=6$
photons have been trapped in each input channel, in the case of the amplitudes of the
control fields at the storing stage such that $\phi^{0}=\pi/8$ and the corresponding zero
phases ($\chi_{2}^{0}=\chi_{3}^{0}$). We show the probability of obtaining exactly 6
photons in the first output channel for the phase $\chi_{3}^{1}=0$ as a function of the
ratio of the amplitudes of the control fields at the release stage (represented by
$\phi^{1}$) and the phase of the first control field $\chi_{2}^{1}$. Apart from 3 peaks
reaching unity, which correspond to the situations when the two channels do not mix at all
or are interchanged, one can see such combinations of the parameters $\phi^{1}$ and
$\chi_{2}^{1}$ that the probability drops to zero. Such a situation can be called
generalized Hong-Ou-Mandel interference and consists in zero probability of the photon
configuration in the output channels identical with that in the input channels. It is
however known that a complete photon coalescence in any of the two output channels can
occur only in the case of storing single photons in both input channels \cite{knight}.
Increasing the number $n$ of photons in the two input channels leads to an increase of the
number of local minima (zeros) of $P(n)$.

\begin{figure}
\includegraphics{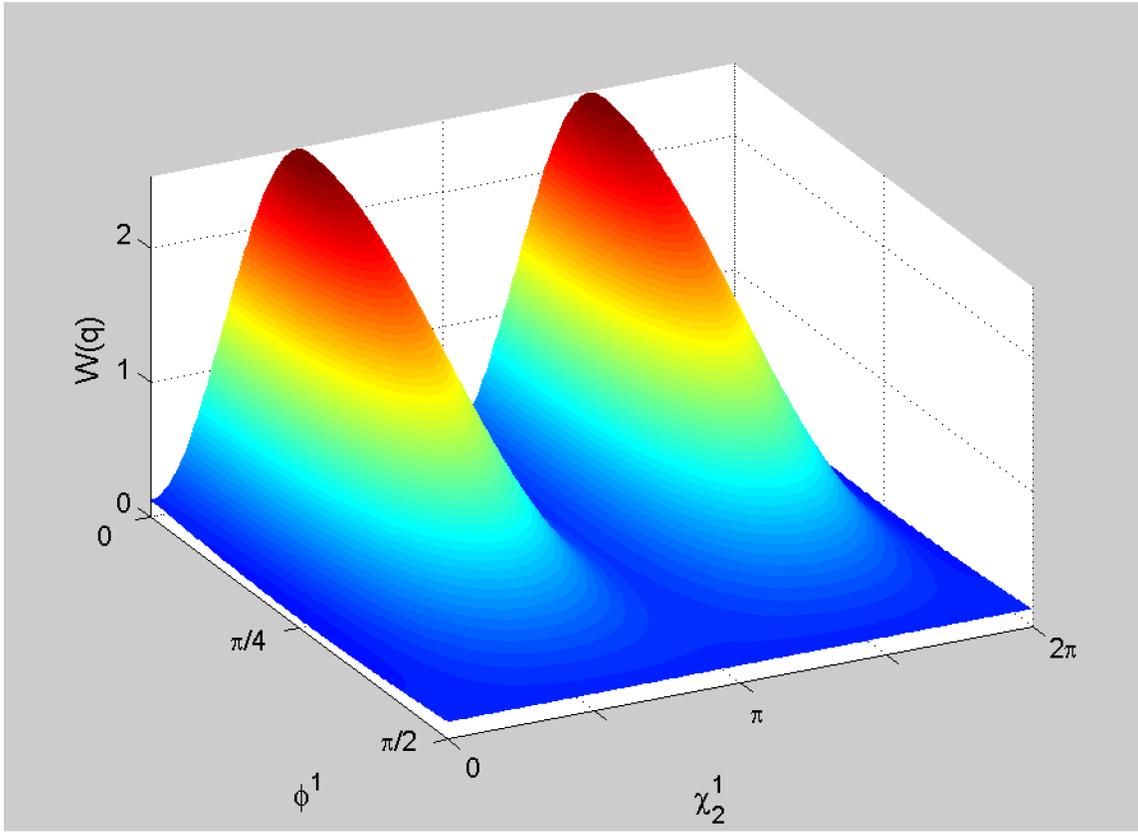}
\caption {\label{fig2} The variance $W(q)$ for the first output channel in the case of
storing squeezed vacuum with $r_{1}=1$, $r_{2}=0.5$, as a function of the angles
$\phi^{1}$ and $\chi_{2}^{1}$ for $\phi^{0}=\pi/4$ and
$\chi_{2}^{0}=\chi_{3}^{0}=\chi_{3}^{1}=0$.}
\end{figure}

\begin{figure}
\includegraphics{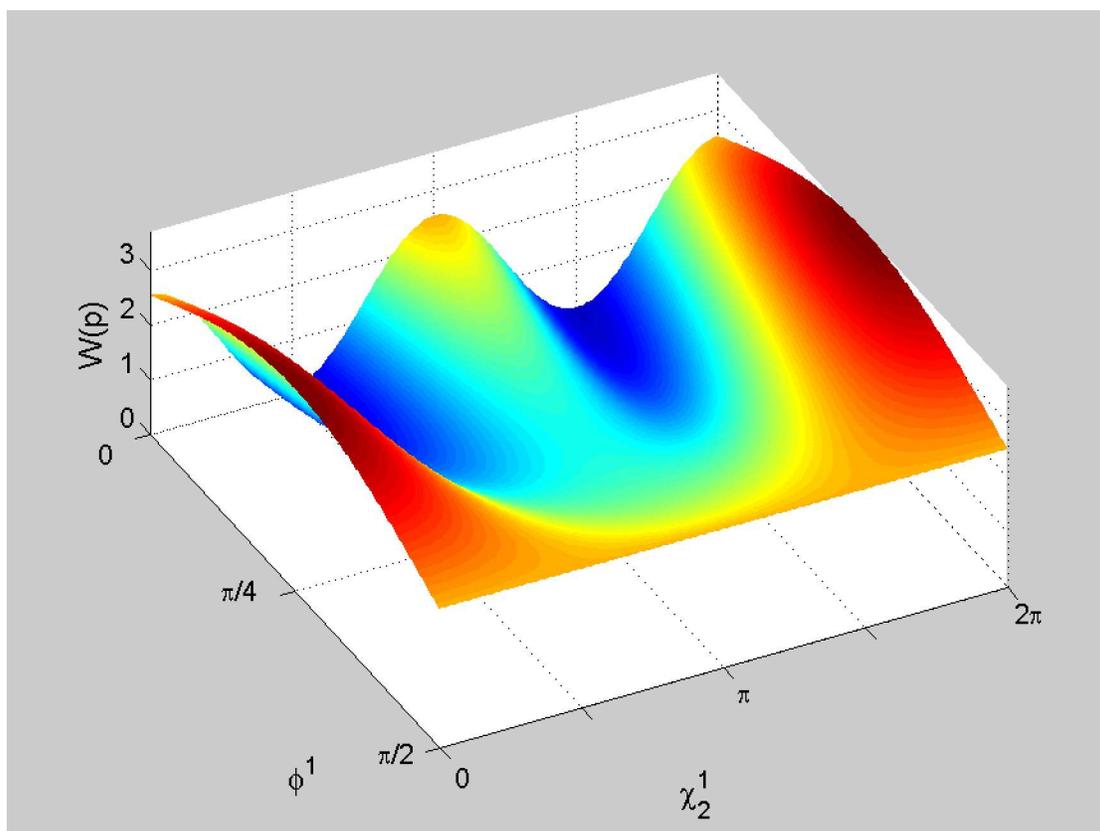}
\caption {\label{fig3} As in Fig. \ref{fig2} but for $W(p)$.}
\end{figure}

\begin{figure}
\includegraphics{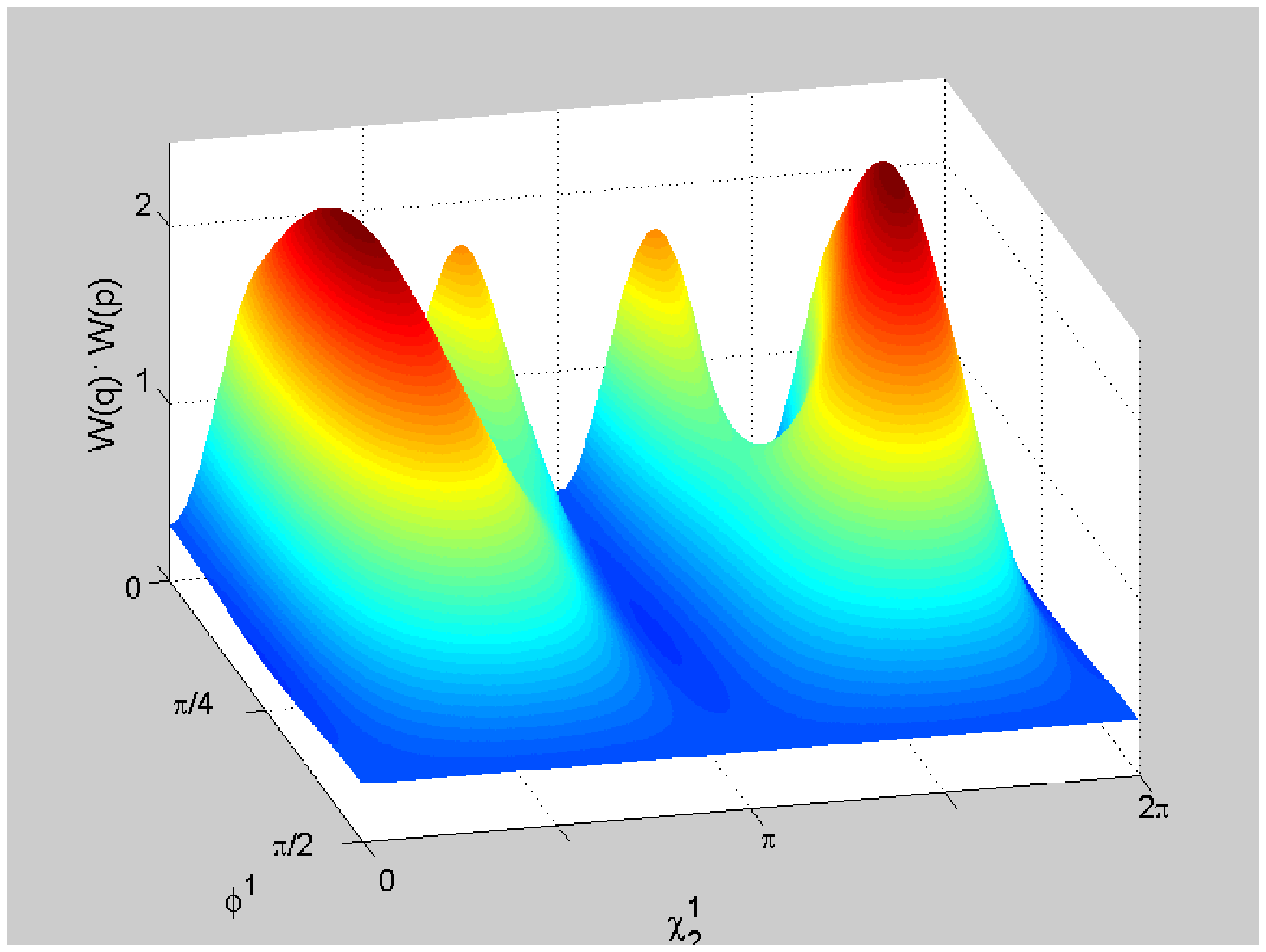}
\caption {\label{fig4} As in Fig. \ref{fig2} but for the product $W(q)W(p)$.}
\end{figure}

Another interesting situation occurs when one stores a quantum squeezed state in each
input channel, assuming overlapping wave packets ($s=1$), which means that after storage
has been performed the medium is in the state $|\alpha_{1}r_{1}>|\alpha_{2}r_{2}>$ (we
consider real squeezing parameters $r_{1,2}$), where
\begin{equation}
|\alpha_{j}r_{j}>=\sum_{n} \frac{\alpha_{j}^{n}}{\sqrt{n!}} A_{j}^{\dagger n}|0>
\exp(-\frac{1}{2}|\alpha_{j}|^{2}),
\end{equation}
where $A_{j}=\cosh r_{j}X^{0}_{j}+\sinh r_{j} X_{j}^{0 \dagger}$. The quadratures
corresponding to the outgoing light in the first output channel are given by
\begin{eqnarray}
q=\frac{1}{\sqrt{2}} (X^{1}_{1}+X^{1\dagger}_{1}),\nonumber\\
p=\frac{-i}{\sqrt{2}} (X^{1}_{1}-X^{1\dagger}_{1}).
\end{eqnarray}
Expressing the operators $X^{1}_{1}$ in terms $X^{0}_{1,2}$ and taking advantage of the
fact that $\alpha_{j}$ is the eigenvalue of $A_{j}$ we obtain the mean values
\begin{eqnarray}
\overline{q}&=&\frac{1}{\sqrt{2}} (u_{1}\alpha_{1}+u_{2}\alpha_{2}) +c.c.\nonumber\\
\overline{q^{2}}&=&\frac{1}{2}[|u_{1}|^{2}(1+2|\alpha_{1}|^{2})+|u_{2}|^{2}(1+2|\alpha_{2}|^{2})]
+\frac{1}{2}[u_{1}^{2}\alpha_{1}^{2}+u_{2}^{2}\alpha_{2}^{2}+2u_{1}u_{2}\alpha_{1}\alpha_{2}+
2u_{1}u_{2}^{*}\alpha_{1}\alpha_{2}^{*}+c.c.],
\end{eqnarray}
where $u_{1}=S_{11}\cosh r_{1}-S_{11}^{*}\sinh r_{1}$, and $u_{2}=S_{12}\cosh
r_{2}-S_{12}^{*}\sinh r_{2}$. Analogous expressions hold for $\overline{p}$ and
$\overline{p^{2}}$ except that $S_{11}$ and $S_{12}$ must be replaced by $-iS_{11}$ and
$-iS_{12}$, respectively. The results are shown in Figs \ref{fig2} and \ref{fig3}.  If
$r_{1}=r_{2}>0$ one can easily show that squeezing in $q$ for the outgoing pulses cannot
be smaller than $\exp(-2r_{1})/2$, which is equal to squeezing in $q$ of the input states,
and cannot be larger than $\exp(2r_{1})/2$, which is squeezing in $p$ of the input states.
One can observe that for real $r_{1,2}$ both incoming pulses minimize the uncertainty
relation which is no longer true for the released pulses. The product of variances of the
quadratures corresponding to the photons in the first output channel is shown in Fig.
\ref{fig4}  as a function of the angles $\phi^{1}$ and $\chi_{2}^{1}$. Though the product
of variances in each input channel is exactly $\frac{1}{4}$, the output values essentially
depend on the mixing properties of the system.

The above-described procedure of storing of two light pulses allows for a realization of a
homodyne-like analysis of the retrieved pulses due to phase-dependent operations on atomic
coherences in which those pulses have been stored. Suppose that one stores photons in a
squeezed state $|\alpha_{1}r_{1}>$ (see Eq. (15)) at the first storage stage and a
Poissonian light in the state $|\alpha_{2}>$ at the second stage. This is a case which for
a symmetric beam splitter constitutes a basis of the balanced homodyne method of
detection. The interesting characteristics of the released light is now the difference
between the number of photons outgoing in the first and second output channels given by
the operator
\begin{eqnarray}
K=X^{1\dagger}_{1}X^{1}_{1}-X^{1\dagger}_{2}X^{1}_{2}=
(|S_{11}|^2-|S_{21}|^{2})X^{0\dagger}_{1}X^{0}_{1}
+(|S_{12}|^2-|S_{22}|^{2})X^{0\dagger}_{2}X^{0}_{2}+\nonumber\\(S_{11}^{*}S_{12}-
S_{21}^{*}S_{22})X^{0\dagger}_{1}X^{0}_{2}+(S_{11}S_{12}^{*}-S_{21}S_{22}^{*})
X^{0\dagger}_{2}X^{0}_{1}.
\end{eqnarray}

Note that the variance of the two last components in the r.h.s of Eq. (18), for the field
2 treated classically (i.e. with the field operators replaced by c-numbers), is
proportional to the variance of the quadrature of the signal field 1 at the angle
determined by the phase of the field 2, the phases of the control fields and/or phases due
to additional interactions. Thus in a symmetric situation (balanced homodyning) in which
$|S_{ij}|^{2}=\frac{1}{2}$ which occurs, e.g., in the case of zero second control field at
the storing stage ($\phi^{0}=0$), equal amplitudes of the control fields at the release
stage ($\phi^{1}=\frac{\pi}{4}$) and zero phases
$\chi^{0}_{2}=\chi^{0}_{3}=\chi^{1}_{3}=0$ we obtain
\begin{equation}
K= \exp(-i\chi^{1}_{2}) X^{0\dagger}_{1}X^{0}_{2} + h.c.
\end{equation}
We again express the operators $X_{1}^{0}$ and $X_{1}{0\dagger}$ in terms of $A_{1}$ and
$A_{1}^{\dagger}$ and take advantage of the facts that the eigenvalues of $A_{1}$ and
$X_{2}$ are, respectively $\alpha_{1}$ and $\alpha_{2}$. That variance of $K$ is for a
classical field 2 of phase $\gamma$
\begin{equation}
W(K)= |\alpha_{2}|^{2} [\cosh 2r_{1}-\sinh2r_{1}\cos2(\chi_{2}^{1}+\gamma)].
\end{equation}
The detection of the squeezed state in the first input channel is thus sensible
equivalently to the phase of the probe classical field in the second input channel (as in
the usual formulation for a beam splitter) as well as to the phase of any of the control
fields. As discussed above, changing the phase of a control field may be equivalent to a
phase shift due to an additional interaction at the storing stage.

\begin{figure}
\includegraphics{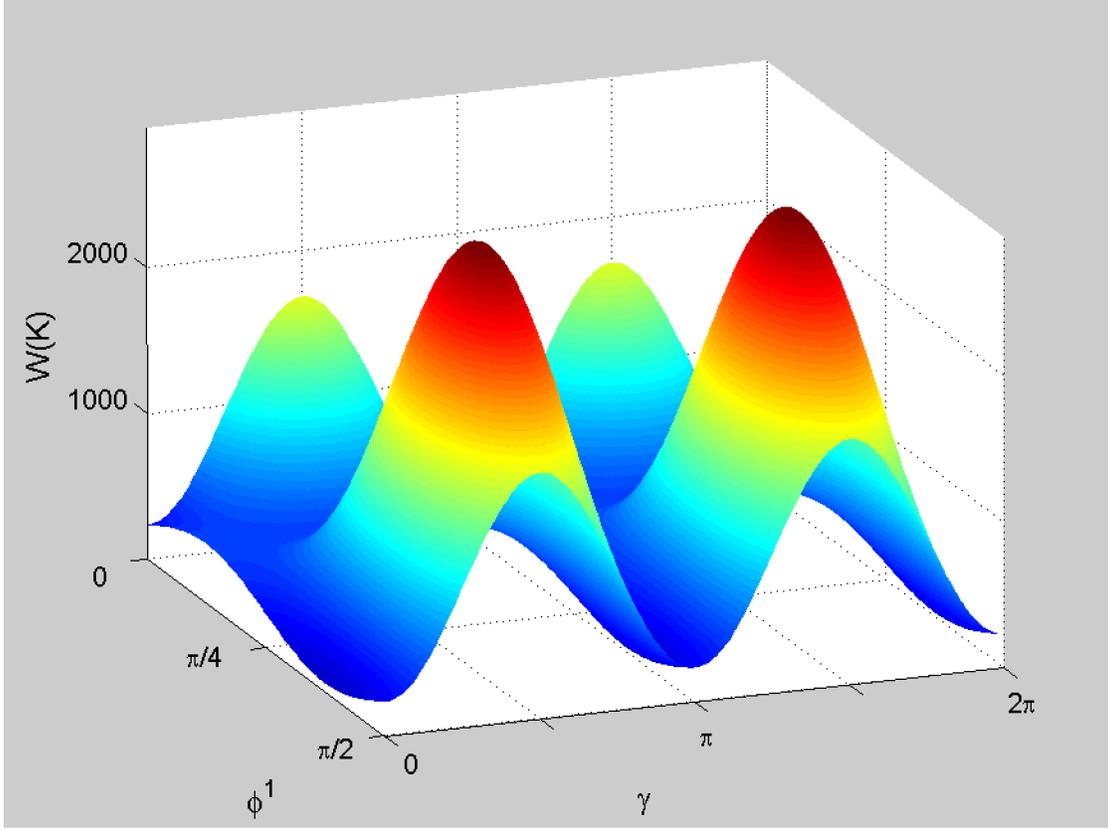}
\caption {\label{fig5} The variance $W(K)$ as a function of the mixing angle $\phi^{1}$
and the phase of the quantum field $\gamma=\arg(\alpha_{2})$ with $|\alpha_{2}|=20$ and
$\phi^{0}=\frac{\pi}{8}$ and $\chi_{i}^{j}=0$.}

\end{figure}

The results are somewhat more complicated in a non-balanced situation. In Fig. \ref{fig5}
we show the variance $W(K)$ as a function of the angle $\phi^{1}$ and the phase $\gamma$
of the Poissonian field in the case of squeezed vacuum in the input channel 1. The
Poissonian field has been treated quantum-mechanically. Then
\begin{equation}
W(K)=\cos^{2}(2\phi^{1}-2\phi^{0})(\frac{1}{2}\sinh^{2}2r_{1}+|\alpha_{2}|^{2})+
\sin^{2}(2\phi^{1}-2\phi^{0})[|\alpha_{2}|^{2}(\cosh2r_{1}-
\sinh2r_{1}\cos2\gamma)+\sinh^{2}r_{1}].
\end{equation}
One can see that for our data the deepest oscillations as a function of $\gamma$ occur for
such amplitudes of the control fields at the release stage that $\phi^{1}=\frac{3\pi}{8}$,
i.e. when $\phi^{1}-\phi^{0}=\frac{\pi}{4}$, so the wave mixing occurs in a symmetric way.
For a given value of $\phi^{1}$ the plot represents two periods of a sine curve with the
amplitude of oscillations and their mean level depending on $\phi^{1}-\phi^{0}$. For
$\phi^{1}=\phi^{0}$ the two input channels do not mix, so there are no oscillations with
changing $\gamma$; then $W(K)$ is the sum of variances of photon numbers in the two input
channels. If the Poissonian field is treated classically the component $\sinh^{2}r_{1}$ in
the last term of Eq. (21) is missing.

%\section{Conclusions}
We have analyzed some statistical properties of the released light, after two quantum
light pulses have been stored in a medium of atoms in the tripod configuration. The
properties of the outgoing pulses have been discussed, depending on the amplitude and
phase relations concerning the control fields. For incoming Fock states we have
investigated the probability distribution of the number of photons released in the two
channels. For the corresponding variances we have also shown how the photon statistics
depends on the overlap of the stored pulses. For stored squeezed states we have shown that
the variances of the quadratures for the released pulses may drastically change depending
on the control fields. We have also given the general formulae for the variance of the
operator of the difference of the photon numbers in the two output channels and discussed
the possibility of a homodyne diagnostics.

\begin{acknowledgments}
The work has been supported by Polish budget funds allocated for the years 2005-2007 as a
research grant No. 1 P03B 010 28. The subject belongs to the scientific program of the
National Laboratory of AMO Physics in Toru\'n, Poland.
\end{acknowledgments}

\newpage

\end{document}